%% file: neurips_2026.tex
\documentclass{article}

\usepackage[dblblindworkshop, final]{neurips_2026}

\usepackage[utf8]{inputenc} %
\usepackage[T1]{fontenc}    %
\usepackage{hyperref}       %
\hypersetup{colorlinks=true,linkcolor=blue!55!black,citecolor=blue!55!black,urlcolor=blue!55!black}
\usepackage{url}            %
\usepackage{booktabs}       %
\usepackage{tabularx}
\usepackage{placeins}
\usepackage{caption}
\usepackage{amsfonts}       %
\usepackage{nicefrac}       %
\usepackage{microtype}      %
\usepackage{xcolor}         %

\usepackage{graphicx}
\usepackage{tikz}
\usetikzlibrary{arrows.meta}
\usepackage{float}
\usepackage{subcaption}
\usepackage{array}
\usepackage{amsmath}
\usepackage{amssymb}
\usepackage[capitalize,noabbrev]{cleveref}

\makeatletter
\renewcommand{\section}{\@startsection{section}{1}{\z@}{-1.2ex \@plus -0.3ex \@minus -0.2ex}{0.7ex \@plus 0.2ex}{\large\bf\raggedright}}
\renewcommand{\subsection}{\@startsection{subsection}{2}{\z@}{-1.0ex \@plus -0.3ex \@minus -0.2ex}{0.4ex \@plus 0.2ex}{\normalsize\bf\raggedright}}
\renewcommand{\@toptitlebar}{\hrule height 4\p@ \vskip 0.15in \vskip -\parskip}
\renewcommand{\@bottomtitlebar}{\vskip 0.18in \vskip -\parskip \hrule height 1\p@ \vskip 0.05in}
\makeatother
\newcommand{\code}[1]{\texttt{#1}}

\title{Harness-Driven Agentic Operation \\ of Optical Pooled Screen Analysis \\[2pt]
  {\large\mdseries An approach to automating and maintaining computational biology pipelines}}
  \workshoptitle{Agentic Life Sciences}
\makeatletter\if@anonymous\renewcommand{\@noticestring}{Submitted to the Agentic Life Sciences
workshop at NeurIPS 2026. Do not distribute.}\fi\makeatother

\author{%
  Matteo Di Bernardo$^{1,2}$ \quad Claire Peterson$^{3}$ \quad Ege Topkoc$^{4,1}$ \\
  Matt Might$^{3}$ \quad Iain M. Cheeseman$^{1,2}$ \\[4pt]
  \normalfont $^{1}$Whitehead Institute for Biomedical Research, Cambridge, MA, USA \\
  \normalfont $^{2}$Department of Biology, Massachusetts Institute of Technology, Cambridge, MA, USA \\
  \normalfont $^{3}$Precision Medicine Institute, University of Alabama at Birmingham, AL, USA \\
  \normalfont $^{4}$Boston University, Boston, MA, USA \\[2pt] \normalfont
  \texttt{matteo.dibernardo@gmail.com}, \texttt{icheese@wi.mit.edu} }

\begin{document}

\maketitle

\begin{abstract}
  Optical pooled screens connect genetic perturbations to changes in cell morphology and behavior,
  but turning the microscopy data into interpretable results takes complex, multi-day analyses.
  Established pipelines process this data, yet researchers still select parameters, inspect outputs,
  and resolve failures for each experiment. Here, we present \emph{Brieflow-auto}, a
  pipeline-specific harness implemented as a Claude Code plugin that lets a coding large language
  model (LLM) operate \emph{Brieflow}, an open-source end-to-end pipeline for optical pooled
  screening data. A researcher supplies experimental facts and approves the launch. An agent tunes
  parameters in an interactive notebook, and the harness validates the configuration, monitors
  execution, recovers from common failures, and checks output quality. For pipeline defects, the
  agent opens upstream issues proposing backward-compatible fixes for the broader community. In five
  previously analyzed screens, harness runs yielded more cells with a confident single-gene
  assignment than the original hand analyses. On a screen the pipeline was not designed for, the
  harness turned each failure into a reviewed upstream fix. More broadly, we propose that wrapping
  an established pipeline in a harness can substantially accelerate its use and the long-term
  reliability of its codebase.

\end{abstract}

\section{Introduction}

Understanding how genes shape cell behavior requires measuring the consequences of genetic
perturbations. Optical pooled screening (OPS) combines pooled genetic perturbations with microscopy,
using an in situ barcode readout to link each cell's appearance to its
perturbation~\citep{feldman2019ops,dibernardo2026brieflow}. Unlike screens read out by sequencing,
OPS requires extracting cellular features from terabytes of images and matching them to perturbation
identities, a multi-day analysis that demands expertise in parameter selection, image inspection,
and troubleshooting.

\begin{figure}[t]
\centering
\resizebox{0.8\linewidth}{!}{\input{figures/overview.tex}}
\caption{Brieflow-auto separates three layers of control. The harness operates a screen repository,
which in turn runs the pipeline. Setup, interview, and approval start a run in which every module
passes the same cycle. The harness recovers from known failures within the run. A pipeline defect becomes an upstream issue, and the reviewed fix returns as a new pipeline version. Blue is agent judgment, white a scripted step,
orange researcher input or review, and grey a record.}
\label{fig:harness}
\end{figure}
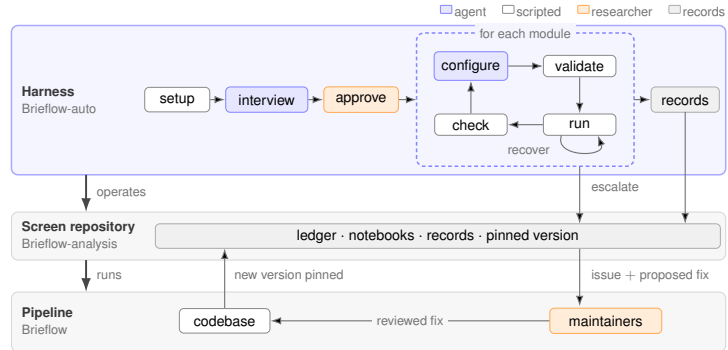

AI agents now support scientific analyses in several ways. Discovery platforms such as Biomni and
Kosmos coordinate analyses across many tools and have led to experimentally validated
discoveries~\citep{huang2025biomni,mitchener2025kosmos}. Coding agents can also apply established
tools under expert guidance. In a recent Anthropic report, Claude autonomously ran de novo protein
binder design campaigns from a single protocol prompt and produced confirmed binders against 14 of
15 testable targets~\mbox{\citep{shanehsazzadeh2026binder}}. Companies such as Tamarind make complex
protein-design tools accessible to researchers as a service with a user interface~\citep{tamarind}.

However, many laboratories and core facilities already have an established analysis pipeline, and
their main cost is running it for each new experiment. For these groups, the analysis itself should
stay fixed while an agent handles configuration, execution, and recovery from failures. This
requires reliable execution and software that can be maintained across experiments, both of which
remain challenging for agents on benchmarks of computational reproduction and iterative
coding~\citep{siegel2024corebench,orlanski2026slopcodebench}. Building a dedicated service for each
pipeline is also a substantial commitment for an academic laboratory or core.

Here, we present \emph{Brieflow-auto}, a harness that lets a coding large language model (LLM)
operate \emph{Brieflow}, an open-source end-to-end pipeline for large-scale OPS microscopy
data~\citep{dibernardo2026brieflow}. The harness encodes our experience with the pipeline and its
failure modes in a Claude Code plugin. It records run state and requires the researcher's approval
before launch, after which the agent handles tuning and most operating decisions without changing
pipeline code. Defects become upstream issues with backward-compatible fixes, so the codebase
evolves through review rather than through edits made during a run. We hypothesize that separating
routine operation from pipeline development reduces manual work while preserving a reproducible,
maintainable codebase.

We describe the changes to Brieflow that enabled agent use, test the harness on five previously
analyzed screens and on a screen outside the pipeline's original use, and release the plugin with an
adaptation guide for other groups' pipelines.

\section{Brieflow}\label{sec:brieflow}

OPS analysis must connect barcode identities with measurements of the same cells, combining
multi-cycle sequencing-by-synthesis (SBS) images with phenotype imaging.
Brieflow~\citep{dibernardo2026brieflow} established an integrated workflow from raw microscope files
through barcode assignment and phenotyping to interpretable groups of perturbations. Running it
still requires configuring each stage, inspecting representative images, and troubleshooting
failures, which limits access for experimental biologists and costs experienced users time.

Brieflow's Snakemake~\citep{snakemake} directed acyclic graph (DAG) provides two properties a
harness needs. Failed jobs resume without repeating completed work, so a tile that runs out of
memory is retried without restarting a multi-day analysis, and the graph is divided into seven
modules (preprocess, SBS, phenotype, merge, classify, aggregate, and cluster) whose boundaries are
checkpoints for quality checks and human intervention. While Brieflow holds the library functions
and module rules, each screen has a separate analysis repository containing its notebooks,
configuration, and a fixed reference to the pipeline version used. This separation lets the harness
configure each screen independently of the shared pipeline code.

Three infrastructure changes made Brieflow practical to run at scale through an agent. We converted
output image formats to chunked, cloud-ready OME-Zarr~\citep{moore2021ngff,zarr} and tables to
Polars~\citep{polars}, removing per-tile read and conversion overhead. In our benchmarks against
tiled TIFF, OME-Zarr gave a median 1.5-fold speedup and 21 percent lower peak memory with concordant
feature outputs, and the advantage widened under concurrency, which made running several screens on
one machine practical. We adapted the Jupyter configuration notebooks, whose hidden execution state
and hard-to-read file changes make them difficult for agents to edit reliably, as
marimo~\citep{marimo} notebooks: plain Python with explicit cell dependencies, so changing a
parameter invalidates only dependent cells and the harness tunes the same notebook the researcher
reads. Finally, we gave the module entry points one shared interface for running a module or the
full sequence.

The largest challenge in OPS analysis is that screens vary in cell density and morphology,
microscope format and metadata, barcode length and chemistry, and phenotype panels. Brieflow exposes
parameters for these differences, and we added a structured experiment description in which the
researcher records them once, before launching an analysis. Together, these changes let the harness
work within a small screen repository with a declared pipeline version, and refer defects found
during a run back to pipeline development.

\section{Brieflow-auto}\label{sec:harness}

Brieflow-auto implements a pipeline-specific harness as a Claude Code
plugin~\citep{claudecodeplugins}. Claude Code supplies the runtime, while Brieflow-auto supplies the
operating instructions, tools, and policies (\cref{fig:harness}). This use of harness follows work
on structured support for long-running agents~\citep{young2025harnesses}. The researcher supplies
experimental facts and approves launch, while the agent inspects data and tunes parameters within
developer-defined ranges. When a required fact is missing, the harness asks the researcher. The
protocol prohibits changing pipeline code or quality thresholds during a run, keeping the scientific
method and acceptance criteria under researcher control. Importantly, a written operating guide
alone would not suffice. In our experience, agent runs failed mainly when the agent lost track of
the run's state rather than for lack of instructions. We therefore implemented each state check,
such as whether a module is complete, whether a value is accepted, and whether a job may resume, as
a script that the harness runs rather than as written guidance.

Brieflow-auto analyzes one screen at a time, and it configures, validates, runs, and checks each of
the screen's modules in turn. Once the agent has selected parameters, the harness checks them
against the pipeline's requirements before launching work, then monitors execution and evaluates the
outputs against established quality criteria. A known operational failure triggers a prescribed,
bounded recovery, such as raising memory and resuming a failed job without repeating completed work.
An unknown failure or a failed quality check pauses for review. Parameters, their sources,
decisions, and completed stages are recorded alongside the analysis, so a run can resume after an
interruption without relying on the agent's earlier conversation.

Although the agent can address some errors and issues as they arise, some problems require changes
to the pipeline itself. For an identified defect, the agent opens an upstream issue with shareable
failure evidence, run context, and a proposed backward-compatible fix. The proposed change must
preserve the existing behavior and capabilities of the overall pipeline, not reshape the pipeline
for one screen. Brieflow's maintainers, the developers who review and merge changes to its code,
then review and release fixes for later analyses, as illustrated in \cref{sec:loop}. The harness
also records each run and, with the researcher's consent, the agent's transcripts, so that proposed
improvements to the harness itself can be tied to evidence from real runs (\cref{app:development}).

\section{Deployment}\label{sec:deploy}

\subsection{Interview and run}

To initiate Brieflow-auto, the researcher first asks the coding agent to configure an analysis and
answers an interview about the experiment. The harness fills parameters from that description,
established defaults, and data inspection, asking only where information is missing. On the two
reanalyses whose interview records survive, 6 and 9 of the 212 resolved variables were questions put
to the researcher, who supplied 14 and 12 values in total (\cref{fig:operator}). The researcher then
reviews the resolved plan and approves the launch. From there, Brieflow-auto enters its run loop and
works through the screen one module at a time. For each module, the agent parametrizes and tunes the
module's notebook, and the harness validates, runs, and checks it before moving to the next. The
harness notifies the researcher as the run progresses and stops only when a failure or a quality
check needs a decision. \Cref{app:interface,app:cycle} illustrate the interview and the run loop.

\begin{figure}[t]
\centering
\includegraphics[width=0.78\linewidth]{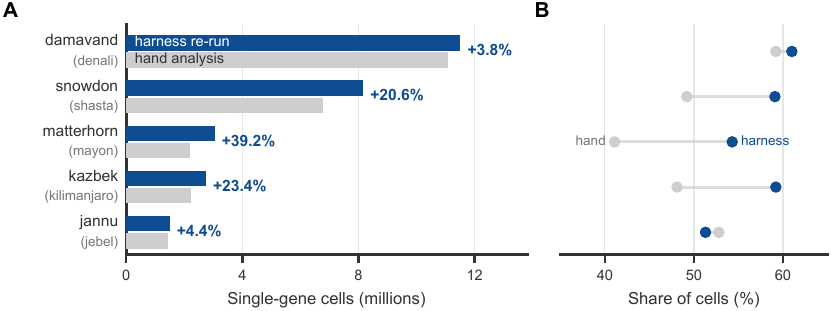}
\caption{Paired reanalysis of five screens, hand analysis against the harness re-run of the same
data. \textbf{A} Cells with a confident single-gene assignment at the end of merge, with the change
for each pair. The count rises in all five pairs. \textbf{B} The same cells as a share of those
entering aggregation, which rises in four of the five pairs (\cref{tab:paired}).}
\label{fig:paired}
\end{figure}

\subsection{Rerunning existing screens}\label{sec:reruns}

To evaluate Brieflow-auto, we reran five previously analyzed screens and compared agent-driven
operation with the original manual analyses, a test of whether the harness preserves analysis
quality while reducing the work of configuring and supervising the pipeline. A critical element of
analyzing OPS data is the tuning of parameters that control barcode assignment, which are inherently
difficult to optimize over one tile of data and require sweeping over a number of individual levers.
With the harness, the agent rapidly sweeps over these parameters in a live notebook kernel and
selects the best combination across a number of tiles in parallel. Compared with the hand analyses,
the harness runs yielded 3.8 to 39.2 percent more cells with a confident single-gene assignment at
the end of merge, in all five pairs and from the same raw images. The agent re-tuned the SBS peak
threshold and the segmentation diameters for each screen. Base calling and barcode assignment used
the same code as the hand analyses. Where both analyses recorded it, the SBS mapping rate also rose,
so the additional reads matched the library rather than adding unmapped noise. The share of cells
entering aggregation that carry such an assignment also increased in four of the five pairs
(\cref{fig:paired}, \cref{tab:paired}). Higher counts measure yield, not assignment accuracy.

\subsection{Extending the pipeline's reach}\label{sec:loop}

After the reanalyses, we tested whether the harness could extend Brieflow to a screen it was not
designed for. We applied Brieflow-auto to a drug screen from another laboratory. As in a knockout
screen, each cell carries an in situ sequenced barcode and is imaged for its phenotype. However, the
barcodes identify constructs rather than knockouts, and each construct is scored under treatment
against its own untreated arm. The screen was also imaged on a microscope whose file format and well
names Brieflow did not yet support, and it used a two-color sequencing chemistry in which one base
carries no dye. When a module failed, the harness classified the failure and reported it with the
error and the context of the run, and each problem led to a pull request with a backward-compatible
fix (\cref{app:maintenance}). For the microscope, the agent identified the metadata and well names
that the pipeline needed to read. For the two-color chemistry, it traced reads that failed to map
back to how bases were called and proposed a base caller suited to the chemistry. For the drug
design, it proposed changes to how constructs are grouped and compared with their untreated arm.
Importantly, these proposals did not always fit the pipeline's code organization, and the Brieflow
maintainers adapted them before merging. Together, these changes extended Brieflow from knockout
screens to drug screens of barcoded constructs within one maintained pipeline.

\section{Conclusions}

Brieflow-auto demonstrates how a coding agent can operate an established pipeline through explicit
configuration, checks, and records. On five different screens, Brieflow-auto yielded more usable
cells than the hand analyses it replaced, and on a screen outside the pipeline's original use each
failure led to a reviewed upstream fix instead of an unreviewed change during the run. This
implementation is narrow, but it addresses a difficult, long-running analysis with many ways to
fail. Many other biological pipelines share these failure modes and the module boundaries that a
harness needs. We therefore expect that a similar harness can be built for them, and we provide an
adaptation guide as a starting point. Broader paired runs with precision checks, time measurements,
and ablations that separate the harness from pipeline changes would test these claims more directly.

\begingroup
\renewcommand{\acksection}{\section*{Author contributions}}
\begin{ack}
M.D. conceived the study, developed Brieflow-auto, performed the analyses, and wrote the manuscript.
C.P. and E.T. developed improvements to Brieflow that enabled the pipeline to be operated by an
agent.\ M.D. carried out all other aspects of the work. M.M. and I.M.C. supervised the project.
I.M.C. edited the manuscript.
\end{ack}
\endgroup

\begin{ack}
We thank Shantanu Singh for feedback on the manuscript and for encouraging this work. We also thank
Eric Lander and Elisa Donnard for encouraging the development of this harness. We thank Molly Moore
and Jason Buenrostro for sharing the drug screen used to extend the pipeline (\cref{sec:loop}) ahead
of its own publication. M.D. was supported by a National Science Foundation Graduate Research
Fellowship (000955563). This work was also supported by the Chan Zuckerberg Initiative (Cell Biology
@ Scale) to I.M.C. Additional support came from Biohub and the Whitehead Innovation Initiative, with
the latter supporting E.T.
\end{ack}

\bibliographystyle{plainnat}
\bibliography{brieflow_auto}

\section*{Data and code availability}
\makeatletter\if@anonymous The harness (plugin and porting guide) and the pipeline are public
repositories. Links are withheld for review. \else The harness is at
\url{https://github.com/cheeseman-lab/brieflow-auto} (plugin, \code{HARNESS.md}). The pipeline is at
\url{https://github.com/cheeseman-lab/brieflow}. \fi\makeatother Both are in active development and
will change after publication. The per-screen tables and figure sources, decision journals, run
manifests, and notebook exports are available on request.

\appendix
\section{Setup and runtime}\label{app:setup}
A setup process transfers the data and prepares the runtime environment on the machine that executes
the analysis, which can be remote or local. We provide a Slurm backend for HPC runs and a local
backend for Google Cloud runs, and these can readily be extended to other compute backends. Claude
Code, the notebook tools, the software environment, and the runner all reside on the execution
machine. The researcher works in a terminal session on that machine, which stays open if the
connection drops. Claude Code Remote Control~\citep{claudecoderemote} lets the researcher follow a
run from another device, and Brieflow-auto sends progress alerts to Slack.

\section{Input data}\label{app:inputs}
The harness takes three inputs, and the researcher supplies all of them before the interview. The
first is the raw imaging data: the multi-cycle SBS images and the phenotype images, staged under one
directory per screen. Setup validates this tree against the layout the pipeline expects, with
sequencing cycles and phenotype rounds under each plate, and converts common alternative layouts. It
reports a missing cycle or plate as a blocker instead of guessing. The second is the barcode library
file, the design table that maps each barcode sequence to its gene. The third is \code{screen.yaml},
a structured description of the experiment that the researcher writes once. It records the
perturbation library (vector, barcode columns and lengths, DNA or RNA detection target, controls),
the SBS acquisition (cycle count, chemistry, channels and base order), the phenotype acquisition
(one entry per channel with its marker), and the cell type and growth conditions.

\FloatBarrier
\section{Interview}\label{app:interface}
\begin{figure}[!htb]
\centering\input{figures/interview.tex}
\caption{Several sources resolve every parameter into one launch review, and only the values no
source can supply become questions for the researcher. The review states each value with its source,
the gaps, and the planned resources before the researcher approves and the run starts. Colors follow
\cref{fig:harness}.}
\label{fig:interview}
\end{figure}
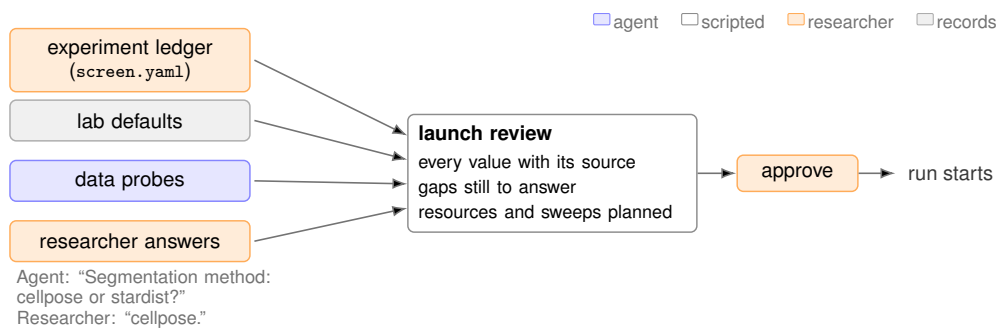

The interview runs once, after the data are staged and setup is complete, and resolves the
parameters of every module before any compute is spent. It separates experimental facts from values
that can be resolved without another question. The agent asks about unresolved facts, then tunes
image-dependent parameters on representative data. Each value is stored with its source
(\cref{fig:interview}).

The run starts only after the researcher approves the launch review.

\FloatBarrier
\section{Run loop}\label{app:cycle}
The harness drives the pipeline's modules through one entry point, \code{bin/brieflow-run}, which
steps through an ordered list of modules and their notebooks. Adding or reordering a module means
editing that list. Flags on the runner choose which modules to run and whether jobs go to a Slurm
cluster or run on the local machine, so the same command works on a cluster, a workstation, or a
cloud virtual machine.

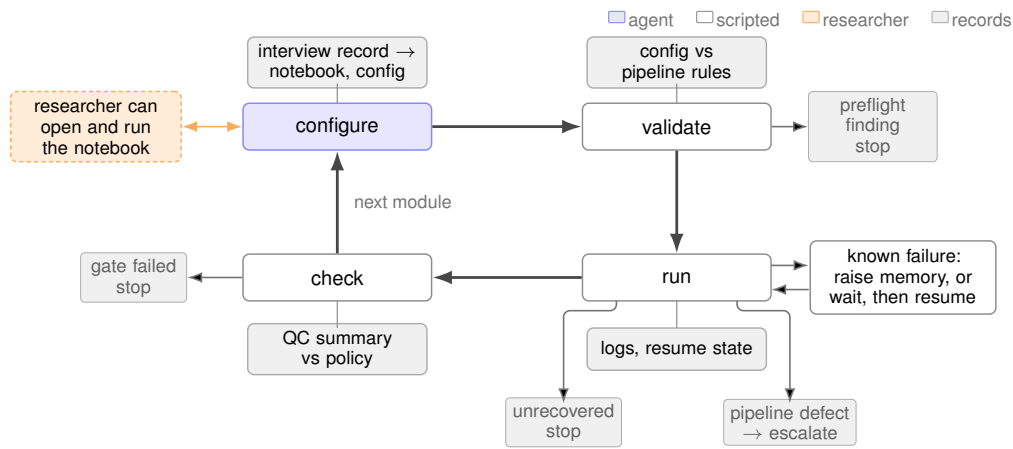
\begin{figure}[!htb]
\centering
\input{figures/loop.tex}
\caption{Every module passes the same cycle, and each step reads and writes a record. The harness
recovers only from known failures, and any other failure stops the run for review. The researcher
can also open and run each module's notebook directly. Colors follow \cref{fig:harness}.}
\label{fig:loop}
\end{figure}

Every module goes through the same four steps, configure, validate, run, and check
(\cref{fig:loop}). First, the agent configures the module. It fills the module's notebook from the
interview and sweeps the tunable parameters in a live marimo kernel, so each trial recomputes only
the cells downstream of the changed value. Second, a preflight check compares the resulting
configuration against the pipeline's code and rejects values that the pipeline would fail on, within
seconds rather than after hours of cluster time. Third, a monitor runs the module. It watches the
jobs, classifies any failure, raises memory and resumes after an out-of-memory error, waits and
resumes after filesystem delays, raises the open-file limit when jobs run out of file handles, and
stops the run for any other failure. Finally, a quality gate reads the module's metrics from the
pipeline's own evaluation outputs and compares them against set thresholds.

The researcher and the agent work in the same notebook. The researcher can open and run any module's
notebook, stop a sweep, change values, and inspect the resulting images, then hand control back once
the changes have been recorded and validated.

\FloatBarrier
\section{Harness details}\label{app:harness}

\paragraph{Records.}\label{app:journal}
Everything the harness resolves or decides is written to files stored with the screen. These include
each parameter and where its value came from (the researcher, the screen description, a lab default,
a data probe, or a sweep), the versions of the harness and pipeline used, and any issues the agent
opened. Important choices also go to a decision journal that marks each as automatic, an agent
judgment, or a researcher decision. For example, a bounded memory increase is recorded as automatic,
a choice between two parameter candidates as agent judgment, and launch approval as a researcher
decision. Another researcher can read these files back to see how a result was produced.

\paragraph{Escalation.}
The runner saves failure evidence locally. The agent investigates pipeline defects and opens GitHub
issues through \code{/brieflow-issue}, following Brieflow's reporting template. Each report includes
a proposed backward-compatible fix. The agent checks for duplicates and removes private data before
submission, and the helper records the issue URL.

\paragraph{Agent boundary.}
The runtime agent can inspect inputs, logs, metrics and images, and set parameters within the ranges
in the defaults file. It can write config and run artifacts, launch, resume, recover or pause
through the provided scripts, and record decisions and findings. It cannot edit pipeline source
during a run, change quality thresholds, guess a fact only the researcher knows, retry an unknown
failure, or publish private run data.

\paragraph{Harness development.}\label{app:development}
The harness records each runner invocation, including its output, before and after state, return
status, and an index of pipeline logs. An opt-in hook also retains the agent's transcripts locally.
These records support a development workflow for the harness itself. GEPA proposes and evaluates
prompt changes from execution feedback, while ACE separates generation, reflection, and curation of
reusable context~\citep{agrawal2025gepa,zhang2025ace}. Our workflow is simpler. The agent selects
local records, describes a problem that can be reproduced and proposes one small change. A script
runs the changed and unchanged harness on the same records, and a change that passes goes to
maintainer review. Evaluating this workflow on independent screens, and adopting a full optimizer
such as GEPA, are promising future directions.

\FloatBarrier
\section{Reanalysis of existing screens}\label{app:deployment}

All five reanalyses ran through every module to clustering, and no quality gate was overridden by
hand.

\paragraph{Usable cells.}
We re-ran five screens previously analyzed by hand (denali, kilimanjaro, mayon, jebel, shasta) end
to end with the harness on cloud virtual machines, producing damavand, kazbek, matterhorn, jannu and
snowdon respectively. We compared the number of usable cells, defined as cells with a confident
single-gene assignment at the end of merge. That count rose in all five pairs from the same raw images, by 3.8, 4.4, 23.4, 39.2 and 20.6 percent (\cref{fig:paired}, \cref{tab:paired}). As a
share of the cells entering aggregation, usable cells rose in four of the five pairs, for example
from 48.1 to 59.2 percent on kazbek and from 41.1 to 54.3 percent on matterhorn. On jannu the share
fell slightly, from 52.8 to 51.3 percent, because 7.9 percent more cells entered merge in the
harness run. The SBS mapping rate, which both analyses recorded for only two pairs, also rose, by
2.4 percentage points on damavand and 11.6 on kazbek. All numbers are read from each run's own
evaluation outputs, so both analyses are measured the same way. The harness also replaced hand-set
round values with measured ones, for example nuclei and cell diameters of 13.97 and 42.14 pixels in
place of 15 and 45 on one screen.

\begin{table}[!htb]
\centering\small
\begin{tabularx}{\linewidth}{@{}Xrrrrr@{}}
\toprule
& \multicolumn{2}{c}{Single-gene cells} & & \multicolumn{2}{c}{Share of cells (\%)} \\
\cmidrule(lr){2-3}\cmidrule(lr){5-6}
Harness / hand pair & Hand & Harness & Change (\%) & Hand & Harness \\
\midrule
damavand / denali    & 11{,}083{,}438 & 11{,}505{,}902 & $+3.8$  & 59.2 & 61.0 \\
jannu / jebel        &  1{,}428{,}007 &  1{,}490{,}127 & $+4.4$  & 52.8 & 51.3 \\
kazbek / kilimanjaro &  2{,}219{,}681 &  2{,}738{,}567 & $+23.4$ & 48.1 & 59.2 \\
matterhorn / mayon   &  2{,}185{,}534 &  3{,}042{,}648 & $+39.2$ & 41.1 & 54.3 \\
snowdon / shasta     &  6{,}764{,}989 &  8{,}156{,}318 & $+20.6$ & 49.2 & 59.1 \\
\bottomrule
\end{tabularx}
\caption{Cells with a confident single-gene assignment at the end of the merge stage, summed over
plates and wells. Counts come from each run's own deduplication summaries and shares are of the
cells entering aggregation, so both analyses are measured the same way. The jannu share fell because more cells entered merge in the harness run.}
\label{tab:paired}
\end{table}

\begin{figure}[!htb]
\centering
\includegraphics[width=\linewidth]{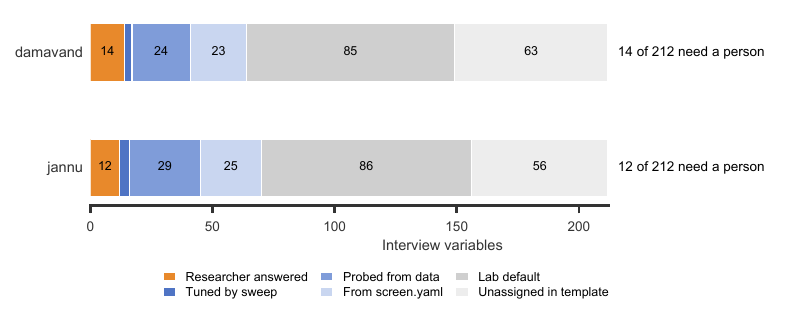}
\caption{Where a screen's parameters come from. All 212 interview variables of a screen, split by
how each value was resolved. Only the orange slice involves a person: values the researcher supplied, of which 6 and 9 were interview questions. Blue values come from parameter sweeps, the data, or the screen description, and ``unassigned in template''
are rows the template carries with no resolved source, such as fixed output paths. The two screens
shown are the reanalyses whose interview records survive.}
\label{fig:operator}
\end{figure}

\paragraph{Researcher effort.}
Each screen's interview resolves 212 notebook parameters. On damavand and jannu, the two reanalyses
whose interview records survive, the researcher was asked 6 and 9 questions and supplied 14 and 12
values in total (\cref{fig:operator}). The remaining values came from the screen description, lab
defaults, data probes, or parameter sweeps. These counts record questions asked, not time spent.

\FloatBarrier
\section{Extending Brieflow to a drug screen}\label{app:maintenance}

\Cref{tab:ledger} lists the changes to Brieflow that came from running the harness on the drug
screen of \cref{sec:loop}, grouped by what each change addressed. Each fix entered the pipeline
itself, so every later screen benefits from it.

\begin{table}[!htb]
\centering\small
\newcommand{\grp}[1]{\multicolumn{3}{@{}l}{\textit{#1}} \\}
\begin{tabularx}{\linewidth}{@{}Xll@{}}
\toprule
Change & Stage & First sign \\
\midrule
\grp{New microscope format}
\#246 Phenix input format & preprocess & missing metadata \\
\#248 per-cycle metadata & preprocess & empty table \\
\#249 Phenix well ids & preprocess & wrong paths$^\dagger$ \\
\#254 HCS column sort & phenotype & job error \\
\#255 well-id passthrough & classify & wrong paths$^\dagger$ \\
\#257 per-well segmentation summary & phenotype & column sums$^\dagger$ \\
\#262 HCS metadata walk & preprocess, phenotype & slow stage \\
\addlinespace
\grp{Sparse fields}
\#250--252 empty-field guards & phenotype & job error \\
\addlinespace
\grp{Two-color chemistry}
\#242 combinatorial base callers & SBS & unmapped reads$^\dagger$ \\
\addlinespace
\grp{Drug-screen design}
\#258 grouped aggregation, features & phenotype, aggregate & wrong reference$^\dagger$ \\
\#293 per-construct untreated reference & aggregate, cluster & wrong reference$^\dagger$ \\
\bottomrule
\end{tabularx}
\caption{Changes to Brieflow from harness runs on the drug screen, grouped by what each change
addressed. All are merged except \#293, which is in review. ``First sign'' is how each problem first
appeared, and $^\dagger$ marks problems that raised no error and were caught by a quality check or an
audit.}
\label{tab:ledger}
\end{table}

\paragraph{Microscope format.}
The extension screen was imaged on an Opera Phenix. Preprocess failed at the first stage, and the
harness classified the failure as a configuration mismatch and reported the failing step and its
error message. The fix was a file-type metadata reader in the pipeline (\#246) and a matching
harness release that probes the format automatically. The resumed run then failed one stage later on
an empty per-cycle metadata table (\#248), then on well identifiers the tile selector could not
split (\#249). Further fixes in the same family followed as later stages encountered the same well
names. One of them raised no error at all: per-well segmentation summaries silently summed each
plate column (\#257), which an audit of the quality outputs caught.

\paragraph{Drug screen.}
The same screen identifies constructs rather than knockouts and compares each construct under
treatment with its untreated arm. Brieflow's aggregation and clustering assume knockouts scored
against non-targeting controls within one treatment, so four parts of the pipeline were extended.
The two-color chemistry needed a combinatorial base caller (\#242). The phenotype stage needed
features defined by the screen, and aggregation needed construct identities grouped by treatment
with the null distribution drawn from the untreated arm (\#258). Scoring then needed each construct
compared with its own untreated arm rather than with a control construct (\#293). Segmentation,
barcode mapping, and merge ran unchanged, and each new parameter entered the screen repository's
notebooks. Every change is reviewed and keeps existing screens working, so Brieflow serves drug
screens of barcoded constructs as well as knockout screens.

\FloatBarrier
\section{Portability}\label{app:portability}
The harness can be adapted to other pipelines that are divided into modules, where each module has
configurable inputs, a way to run it, a signal that it has finished, a way to resume after a
failure, and outputs that can be checked for quality. Brieflow and Snakemake provide these here.

The porting guide released with the harness (\code{HARNESS.md}) divides the harness code into three
parts. The first can be reused without changes: the list of modules, the loop that configures,
validates, runs, and checks each module, the job monitor, and the connections to each compute
scheduler. The second must be rewritten for a new pipeline: how a module is configured (here, a
marimo notebook), how failures are read from logs, and the checks before and after each module runs.
The third is policy that each group sets from its own experience: how much to raise memory after a
failure, how long to wait on a slow filesystem, and the quality thresholds that define a passing
result.

\end{document}

%% file: figures/overview.tex
\input{figures/schematic_style.tex}%
\resizebox{\linewidth}{!}{%
\begin{tikzpicture}[x=1cm,y=1cm]
\hyphenpenalty=10000 \exhyphenpenalty=10000
\schcanvas{6.80}

\draw[schbandmain] (0.30,3.62) rectangle (13.70,6.42);
\draw[schband]     (0.30,2.02) rectangle (13.70,2.92);
\draw[schband]     (0.30,0.30) rectangle (13.70,1.42);

\node[schlabel,anchor=west,align=left] at (0.42,5.03)
  {\schbandtitle{Harness}{Brieflow-auto}};
\node[schlabel,anchor=west,align=left] at (0.42,2.47)
  {\schbandtitle{Screen repository}{Brieflow-analysis}};
\node[schlabel,anchor=west,align=left] at (0.42,0.88)
  {\schbandtitle{Pipeline}{Brieflow}};

\schlegend{13.70}{6.68}

\draw[schloop] (1.70,3.62) -- (1.70,2.92);
\node[schnote,anchor=west] at (1.86,3.27) {operates};
\draw[schloop] (1.70,2.02) -- (1.70,1.42);
\node[schnote,anchor=west] at (1.86,1.72) {runs};

\node[schstep,minimum width=1.22cm] (setup) at (3.41,5.03) {setup};
\node[schstep,schagent,minimum width=1.54cm] (intv) at (5.09,5.03) {interview};
\node[schstep,schhuman,minimum width=1.40cm] (appr) at (6.86,5.03) {approve};
\node[schstep,schrecord,minimum width=1.28cm] (recs) at (12.94,5.03) {records};

\draw[schsleeve] (7.90,3.78) rectangle (11.98,6.28);
\node[schnote,fill=blue!4,inner sep=2.5pt] at (9.94,6.28) {for each module};

\draw[schflow] (setup) -- (intv);
\draw[schflow] (intv) -- (appr);
\draw[schflow] (appr.east) -- (7.90,5.03);
\draw[schflow] (11.98,5.03) -- (recs.west);

\node[schring,schagent] (cfg) at (8.92,5.66) {configure};
\node[schring]          (val) at (10.96,5.66) {validate};
\node[schring]          (run) at (10.96,4.56) {run};
\node[schring]          (chk) at (8.92,4.56) {check};

\draw[schflow] (cfg) -- (val);
\draw[schflow] (val) -- (run);
\draw[schflow] (run) -- (chk);
\draw[schflow] (chk) -- (cfg);

\draw[schflow] (10.62,4.36) .. controls (10.46,3.94) and (11.46,3.94) .. (11.30,4.36);
\node[schnote,anchor=east] at (10.44,4.08) {recover};

\draw[schflow] (12.94,4.80) -- (12.94,2.70);

\draw[schflow] (10.96,3.78) -- (10.96,2.70);
\draw[schflow] (10.96,2.24) -- (10.96,1.15);
\node[schnote,anchor=west] at (11.14,3.40) {escalate};
\node[schnote,anchor=west] at (11.14,1.72) {issue $+$ proposed fix};

\node[schrecord,minimum height=0.46cm,minimum width=10.60cm,inner sep=3pt] (sr) at (8.30,2.47)
  {ledger $\cdot$ notebooks $\cdot$ records $\cdot$ pinned version};

\node[schstep,minimum width=1.70cm] (code) at (4.30,0.88) {codebase};
\node[schstep,schhuman,minimum width=2.10cm] (mnt) at (11.45,0.88) {maintainers};

\draw[schflow] (mnt.west) -- (code.east);
\node[schnote,fill=schBandFill,inner sep=2.5pt] at (7.78,0.88) {reviewed fix};

\draw[schflow] (code.north) -- (4.30,2.24);
\node[schnote,anchor=west] at (4.44,1.72) {new version pinned};

\end{tikzpicture}}

%% file: figures/interview.tex
\input{figures/schematic_style.tex}%
\resizebox{\linewidth}{!}{%
\begin{tikzpicture}[x=1cm,y=1cm]
\hyphenpenalty=10000 \exhyphenpenalty=10000
\schcanvas{4.45}

\schlegend{13.70}{4.35}

\tikzset{schsrc/.style={schstep,text width=2.90cm,minimum height=0.54cm}}
\node[schsrc,schhuman]  (led) at (2.20,3.85) {experiment ledger ({\schMono screen.yaml})};
\node[schsrc,schrecord] (def) at (2.20,3.05) {lab defaults};
\node[schsrc,schagent]  (prb) at (2.20,2.25) {data probes};
\node[schsrc,schhuman]  (ans) at (2.20,1.45) {researcher answers};

\node[schnote,anchor=north west,align=left,text width=4.10cm] at (0.66,1.12)
  {Agent: ``Segmentation method: cellpose or stardist?''\\
   Researcher: ``cellpose.''};

\node[schbox,align=left,text width=3.55cm,minimum height=1.35cm] (lr) at (7.80,2.35)
  {\textbf{launch review}\\[2pt]
   {\schTiny every value with its source}\\[0.5pt]
   {\schTiny gaps still to answer}\\[0.5pt]
   {\schTiny resources and sweeps planned}};

\draw[schflow] (3.85,3.85) -- (5.88,2.85);
\draw[schflow] (3.85,3.05) -- (5.88,2.53);
\draw[schflow] (3.85,2.25) -- (5.88,2.21);
\draw[schflow] (3.85,1.45) -- (5.88,1.89);

\node[schstep,schhuman,minimum width=1.60cm] (app) at (11.05,2.35) {approve};
\draw[schflow] (lr.east) -- (app.west);
\draw[schflow] (app.east) -- (12.35,2.35);
\node[schlabel,anchor=west] at (12.44,2.35) {run starts};

\end{tikzpicture}}

%% file: figures/loop.tex
\input{figures/schematic_style.tex}%
\resizebox{\linewidth}{!}{%
\begin{tikzpicture}[x=1cm,y=1cm]
\hyphenpenalty=10000 \exhyphenpenalty=10000
\schcanvas{6.30}

\schlegend{13.70}{6.18}

\tikzset{schfour/.style={schstep,minimum width=2.50cm,minimum height=0.62cm}}
\node[schfour,schagent] (cfg) at (4.75,4.77) {configure};
\node[schfour]          (val) at (9.25,4.77) {validate};
\node[schfour]          (run) at (9.25,2.77) {run};
\node[schfour]          (chk) at (4.75,2.77) {check};

\draw[schloop] (cfg) -- (val);
\draw[schloop] (val) -- (run);
\draw[schloop] (run) -- (chk);
\draw[schloop] (chk) -- (cfg);
\node[schnote,anchor=west] at (4.93,3.78) {next module};

\tikzset{schrec/.style={schrecord,font=\schTiny,text width=2.20cm,
  minimum height=0.56cm,inner sep=2.5pt}}
\node[schrec] at (4.75,5.62) {interview record $\rightarrow$ notebook, config};
\node[schrec] at (9.25,5.62) {config vs pipeline rules};
\node[schrec] at (9.25,1.82) {logs, resume state};
\node[schrec] at (4.75,1.82) {QC summary vs policy};

\draw[schlink] (4.75,5.08) -- (4.75,5.34);
\draw[schlink] (9.25,5.08) -- (9.25,5.34);
\draw[schlink] (9.25,2.46) -- (9.25,2.10);
\draw[schlink] (4.75,2.46) -- (4.75,2.10);

\node[schhuman,dash pattern=on 1.6pt off 1.2pt,font=\schTiny,text width=2.05cm,
  minimum height=0.62cm,inner sep=3pt] (op) at (1.55,4.77)
  {researcher can open and run the notebook};
\draw[{Latex[length=2.0mm,width=1.45mm]}-{Latex[length=2.0mm,width=1.45mm]},
  line width=0.6pt,draw=schHumanLine] (op.east) -- (cfg.west);

\node[schbox,font=\schTiny,text width=2.25cm,minimum height=0.74cm,inner sep=3pt]
  (rec) at (12.25,2.77) {known failure: raise memory, or wait, then resume};
\draw[schflow] (10.50,2.93) -- (11.00,2.93);
\draw[schflow] (11.00,2.61) -- (10.50,2.61);

\node[schexit,text width=1.55cm] (e6) at (11.85,4.77) {preflight finding\\stop};
\draw[schflow] (val.east) -- (e6.west);

\node[schexit,text width=1.25cm] (e1) at (2.05,2.77) {gate failed\\stop};
\draw[schflow] (chk.west) -- (e1.east);

\node[schexit,text width=1.38cm] (e2)  at (7.75,0.85) {unrecovered\\stop};
\node[schexit,text width=1.60cm] (esc) at (10.75,0.85) {pipeline defect\\$\rightarrow$ escalate};
\draw[schflow,rounded corners=2.4pt] (8.45,2.46) -- (8.45,2.33) -- (7.75,2.33) -- (7.75,1.10);
\draw[schflow,rounded corners=2.4pt] (10.05,2.46) -- (10.05,2.33) -- (10.75,2.33) -- (10.75,1.10);

\end{tikzpicture}}